\documentclass{eas}
\usepackage{graphicx}
\begin{document}

%%-----------------------------
%%      the top matter
%%-----------------------------
\title{Atmospheric dynamics of red supergiant stars and interferometry} 
\author{Chiavassa, A.}\address{GRAAL, cc072, Universit\'e Montpellier II, F-34095 Montpellier cedex 05, France}
%\author{...}\address{...}
%\author{...}\address{...}
%
%
\begin{abstract}

We developed a 3D pure LTE radiative transfer code to derive observables expected for RSGs, with emphasis on small scale structures, from radiative-hydrodynamic (RHD)  simulations of red supergiant stars (RSGs) carried out with CO5BOLD (\cite{2002AN....323..213F}). We show that the convection-related surface structures are observable with today's interferometers. Moreover, the RHD simulations are a great improvement over parametric models for the interpretation of interferometric observations.

\end{abstract}
\maketitle
%%-----------------------------
%%      your text
%%-----------------------------

\section{Introduction}
Red supergiant (RSG) stars are irregular, small-amplitude variables and they have a very complex spectrum where strong molecular absorption by TiO and other molecules dominates. This leads to an ill-defined continuum. Furthermore, \cite{2007A&A...469..671J} find that high-resolution spectroscopic time-series observations of a sample of RSG reveal that the variable velocity field in the atmosphere probably has convective origin. The velocities are supersonic and vary with a time-scale of a few $100$ days and the resulting spectral lines have a strong broadening together with asymmetries and wavelength shifts. For all these reasons, spectral synthesis of these objets is really difficult to achieve. \\
The observed strong line asymmetries indicate that convection consists of giant cells, as suggested by 
radiative hydrodynamic modeling (see Freytag contribution in this volume).

\subsection{Radiative-hydrodynamic simulations} 

The numerical simulations analyzed here are performed with
``CO5BOLD'' (``COnservative COde for the COmputation of COmpressible COnvection
in a BOx of L Dimensions, l=2,3'') developed by Freytag, Steffen, Ludwig and collaborators (see \cite{2002AN....323..213F}).
The RSG simulations presented employ the \emph{global star-in-a-box} setup:
the computational domain is a cube, and the grid is equidistant in all directions. Radiation transport is strictly LTE. The grey Rosseland mean opacity is a function of pressure and temperature. The necessary values are found by interpolation in a 2D table which
has been merged at around 12\,000\,K from
high-temperature OPAL data (\cite{1992ApJ...397..717I}) and low-temperature PHOENIX data (\cite{1997ApJ...483..390H}) by Hans-G{\"u}nter Ludwig.\\
Some more technical informations can be found in the CO5BOLD Online User Manual and in the contribution by Freytag (this volume).

\subsection{A posteriori radiative transfer in the hydrodynamical simulations} \label{3dcode}

\begin{figure*}
\centering
\includegraphics[angle=0,width=0.6\hsize]{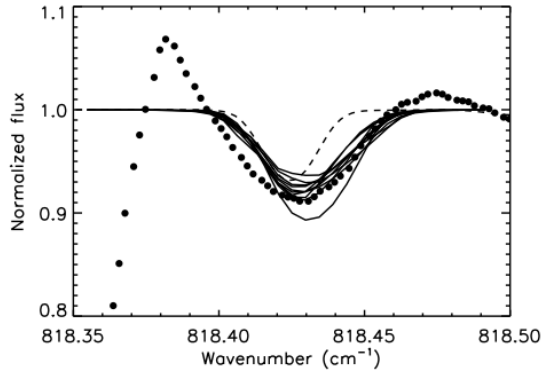} 
\caption{The time sequence computed for the RHD model described in the text (time-step of about $23$ days) is compared to  observations (dots; \cite{2006ApJ...637.1040R}; the sharp decrease below $818.38$cm$^{-1}$ is an instrumental artifact). The dashed line is a one-dimensional hydrostatic MARCS model calculation convolved with an exponential profile of $13$km/s width.}
\label{spec}
\end{figure*}

We performed 3D pure LTE radiative transfer calculations in snapshots of the 3D hydrodynamical simulations (the RHD model presented in this work has a solar composition, $235^3$ grid points, M = $12M_{\odot}$, L = $100000L_{\odot}$, T$_{eff}$ = $3482$K, $\log g$ = $-0.39$, and R = $880R_{\odot}$), taking into account the Doppler shifts caused by convective motions. The emerging intensity at the top of one ray of the RHD computational box, is computed by summing the contribution of the source function at different depths.\\
To reduce the computing time, the extinction coefficients per unit mass are pre-tabulated as a function of temperature, density and wavelength with a resolution of $0.01$ \AA \ using MARCS (\cite{1992A&A...256..551P}, \cite{2003ASPC..288..331G}) opacity data. We checked that this resolution is sufficient to ensure an accurate characterization of line profiles even after interpolation at the Doppler shifted wavelengths (\cite{2006sf2a.conf..455C}).\\
The a posteriori radiation transfer treatment is used to compute : (i) spectra at high and low (Opacity Sampling) spectral resolution that are compared with spectroscopic and spectrophotometry observations; (ii) intensity maps to compare with interferometric observations.\\
Concerning the spectral synthesis, Fig. \ref{spec} shows that RHD models approximately reproduce the $12\mu m$ H$_2$O line width without the need for micro- or macrotubulence  (which is needed in the one-dimensional model). The shifts of the calculated line center span values between -1 and -3 km/s with respect to the hydrostatic model. The profiles are asymmetric. 

\section{Interferometric imaging of red supergiant stars}

\begin{figure*}
\centering
\begin{tabular}{cc}
\includegraphics[angle=0,width=0.33\hsize]{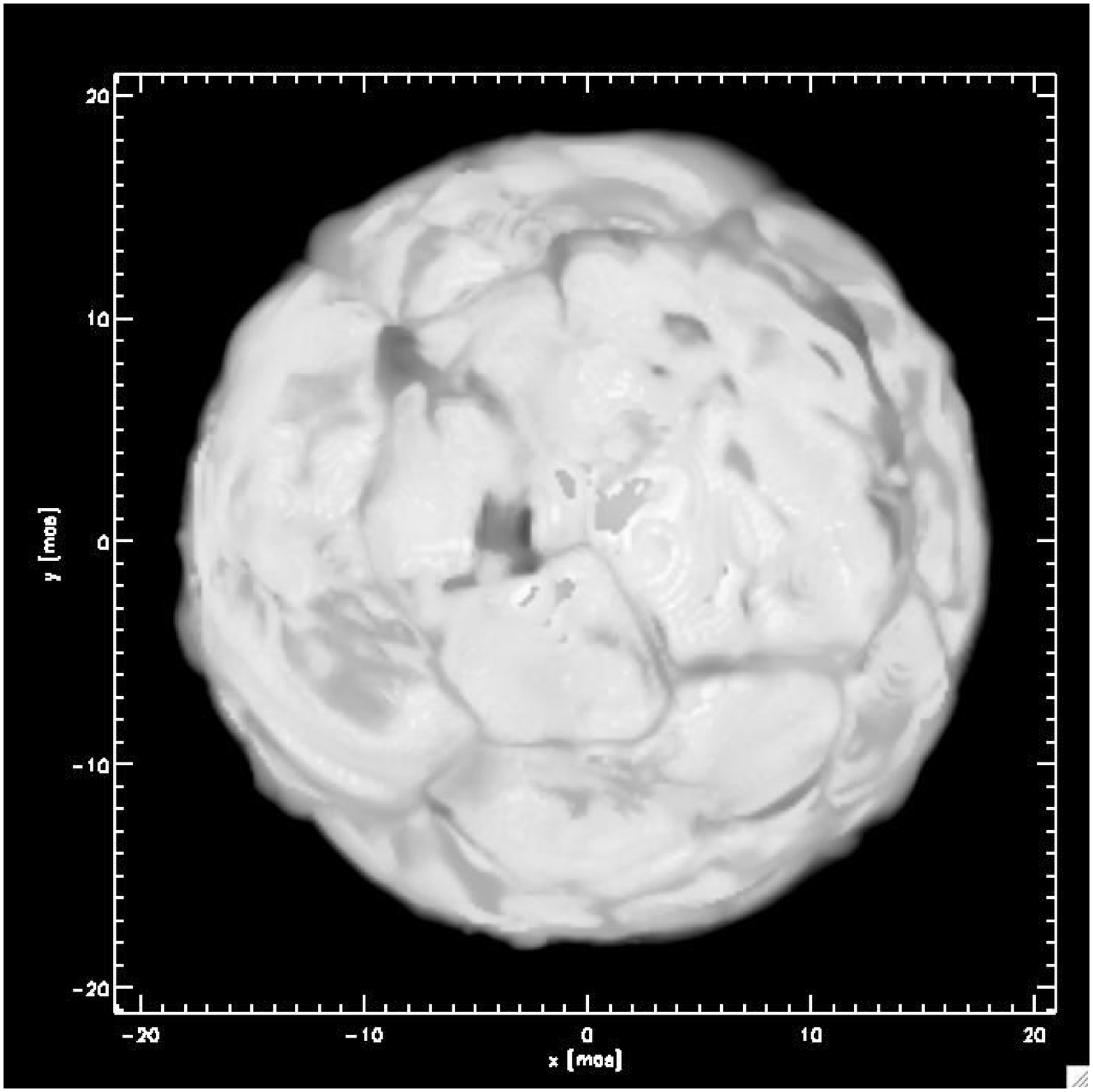} 
\includegraphics[angle=0,width=0.49\hsize]{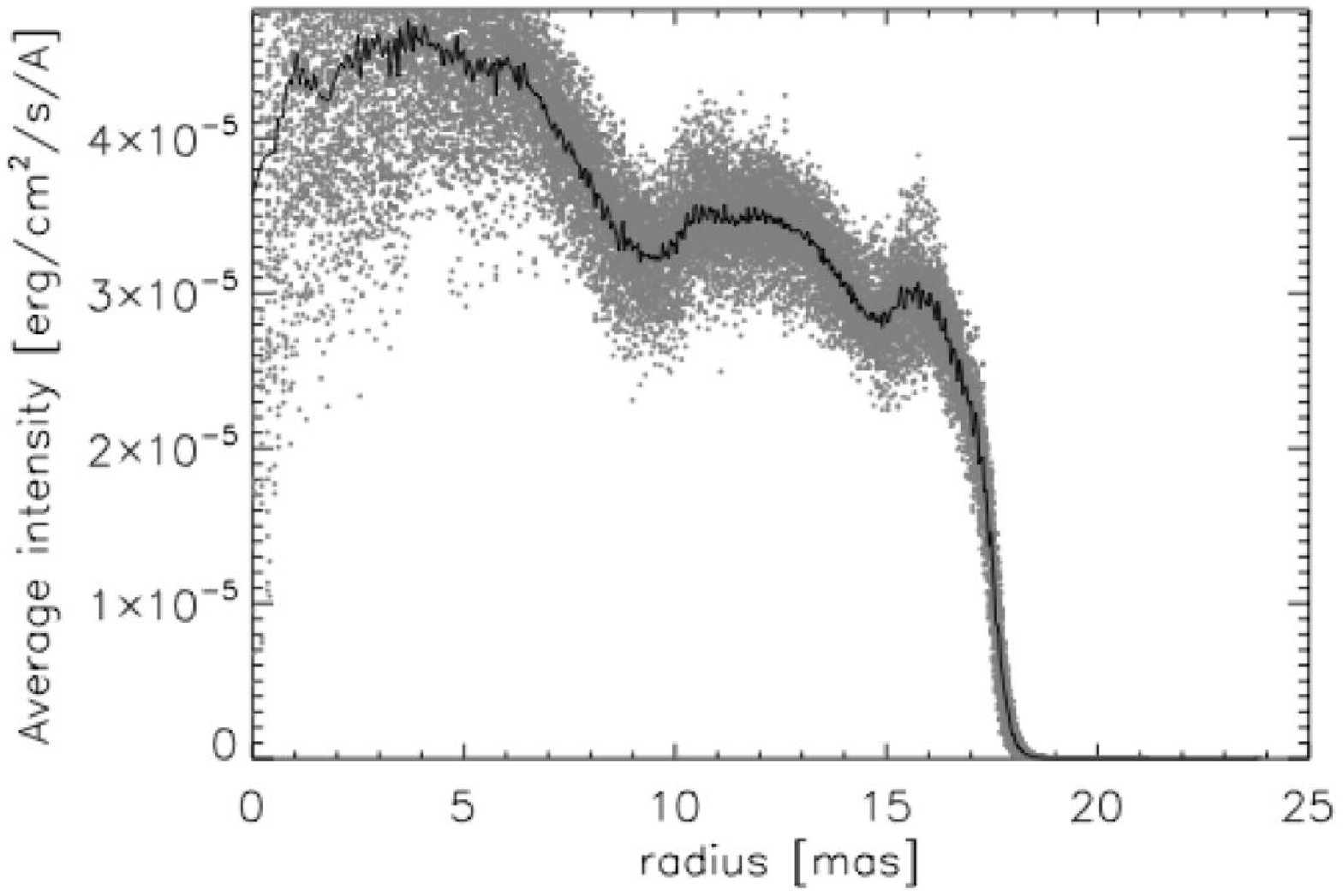} \\
\includegraphics[angle=0,width=0.33\hsize]{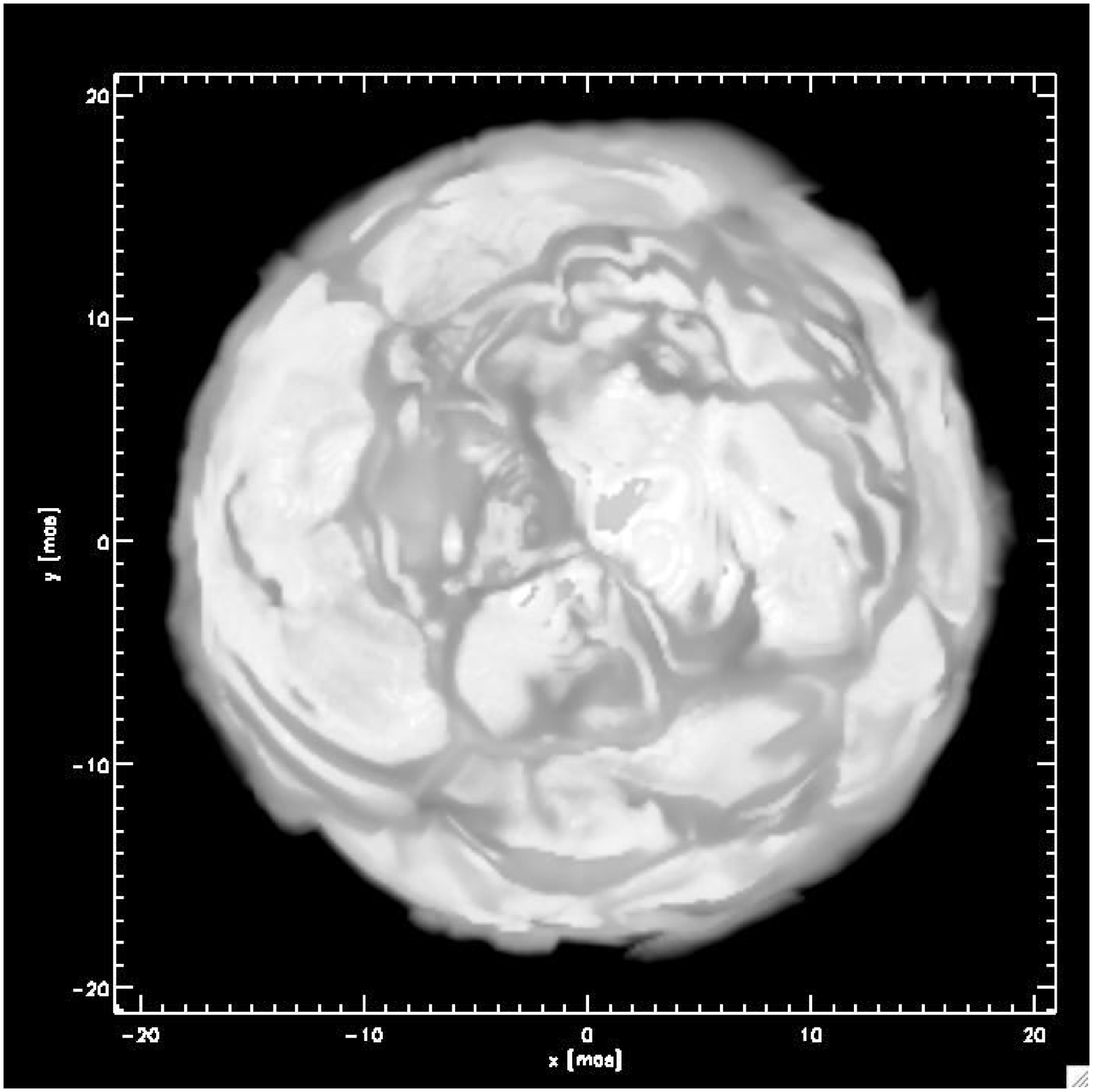} 
\includegraphics[angle=0,width=0.49\hsize]{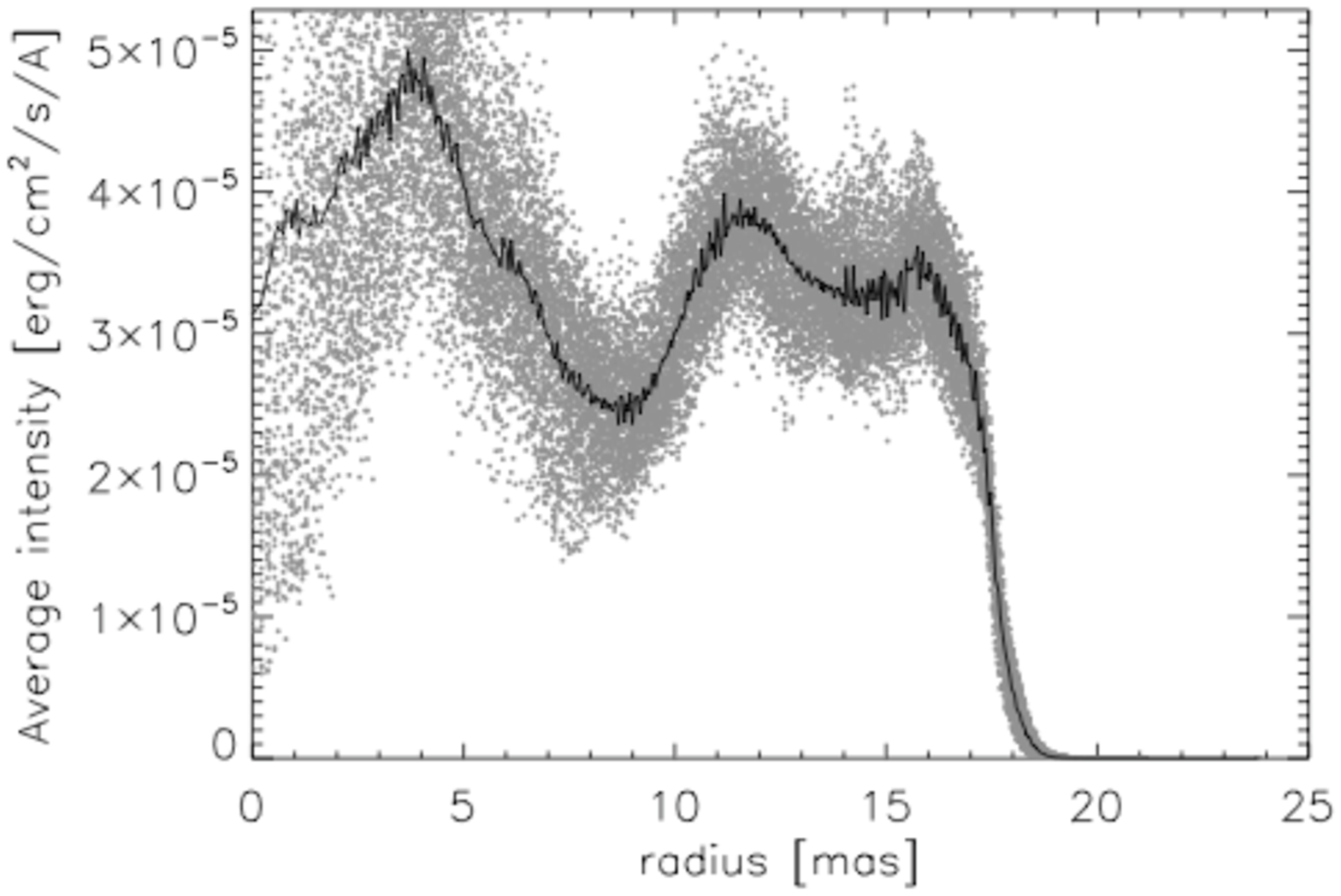} \\
\end{tabular}
\caption{Logarithmic maps of monochromatic intensity (a grey scale is used and lighter shades correspond to higher intensities) and scatter-plot of the radially averaged intensity computed for $48$ snapshots from RHD simulations described in the text with a time-step of about $23$ days (grey; the black line is the averaged profile). \emph{Upper row} : continuum at $1.6\mu m$ ($H^-$ minimum opacity). \emph{Bottom row} : Nearby CO line.}
\label{intensity}
\end{figure*}

The characterization of stars beyond simple limb-darkened circular disks is a very important task to be achieved. Our 3D radiative transfer code is employed to derive observables expected for RSGs with emphasis to small scale structures. We produced a time-sequence, with a time-step of about $23$ days covering about one stellar year, of maps of monochromatic intensity at two different wavelengths. These wavelengths probe different depths in the stellar atmosphere since they correspond to the continuum at $1.6\mu m$, where the $H^-$ minimum opacity occurs and one can see deeper in the photosphere, and a nearby CO line. Two examples of the maps are displayed in Fig. \ref{intensity}, together with the radially averaged intensity distribution. Absolute model dimensions have been scaled to approximately match the angular diameter of $35$mas. The contrast between large bright granules visible in the continuum (upper-left panel), and the smaller structures seen in the CO line (bottom-left panel) is striking.

\begin{figure*}
\centering
\begin{tabular}{cc}
\includegraphics[angle=0,width=0.48\hsize]{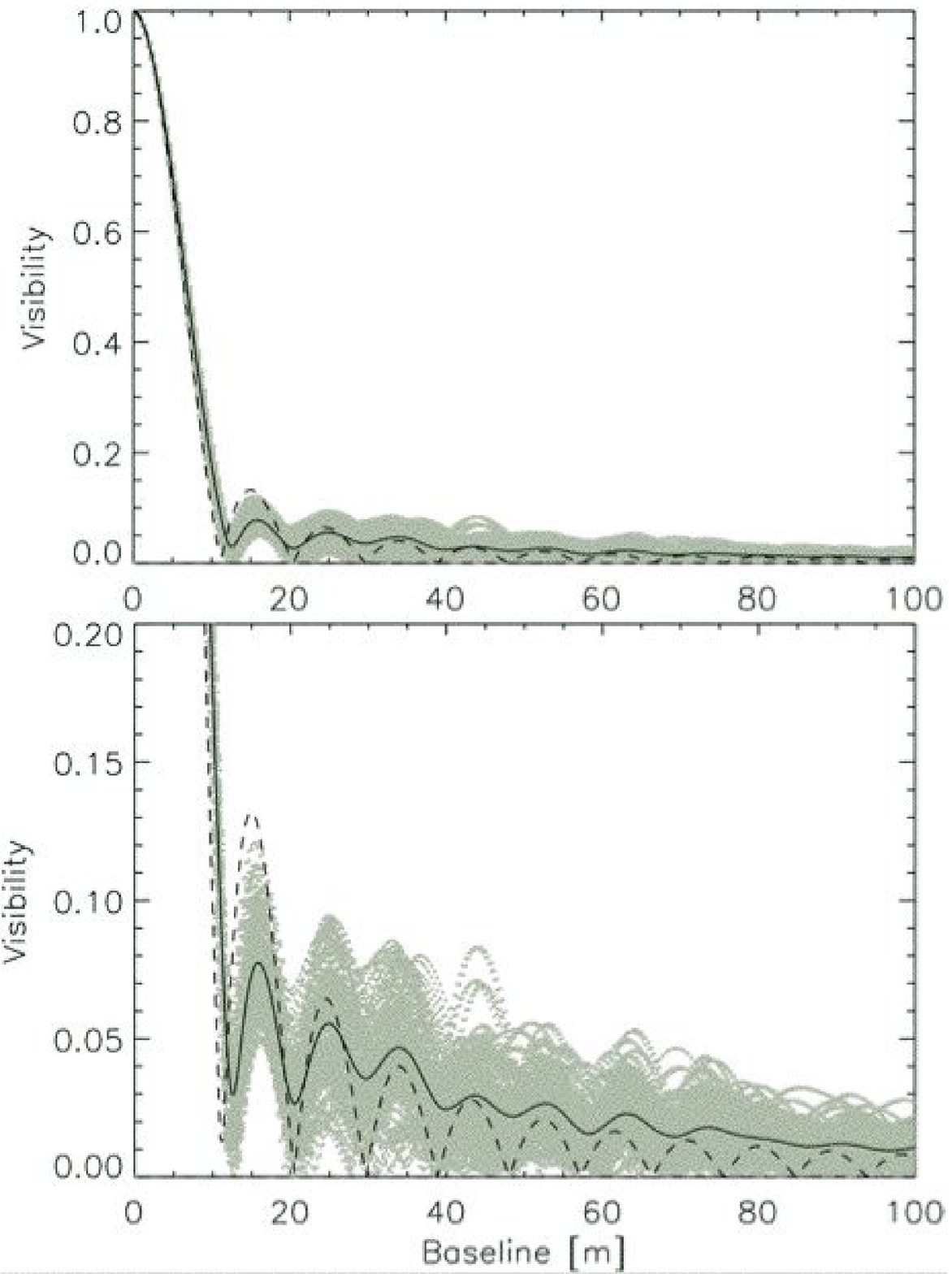} 
\includegraphics[angle=0,width=0.48\hsize]{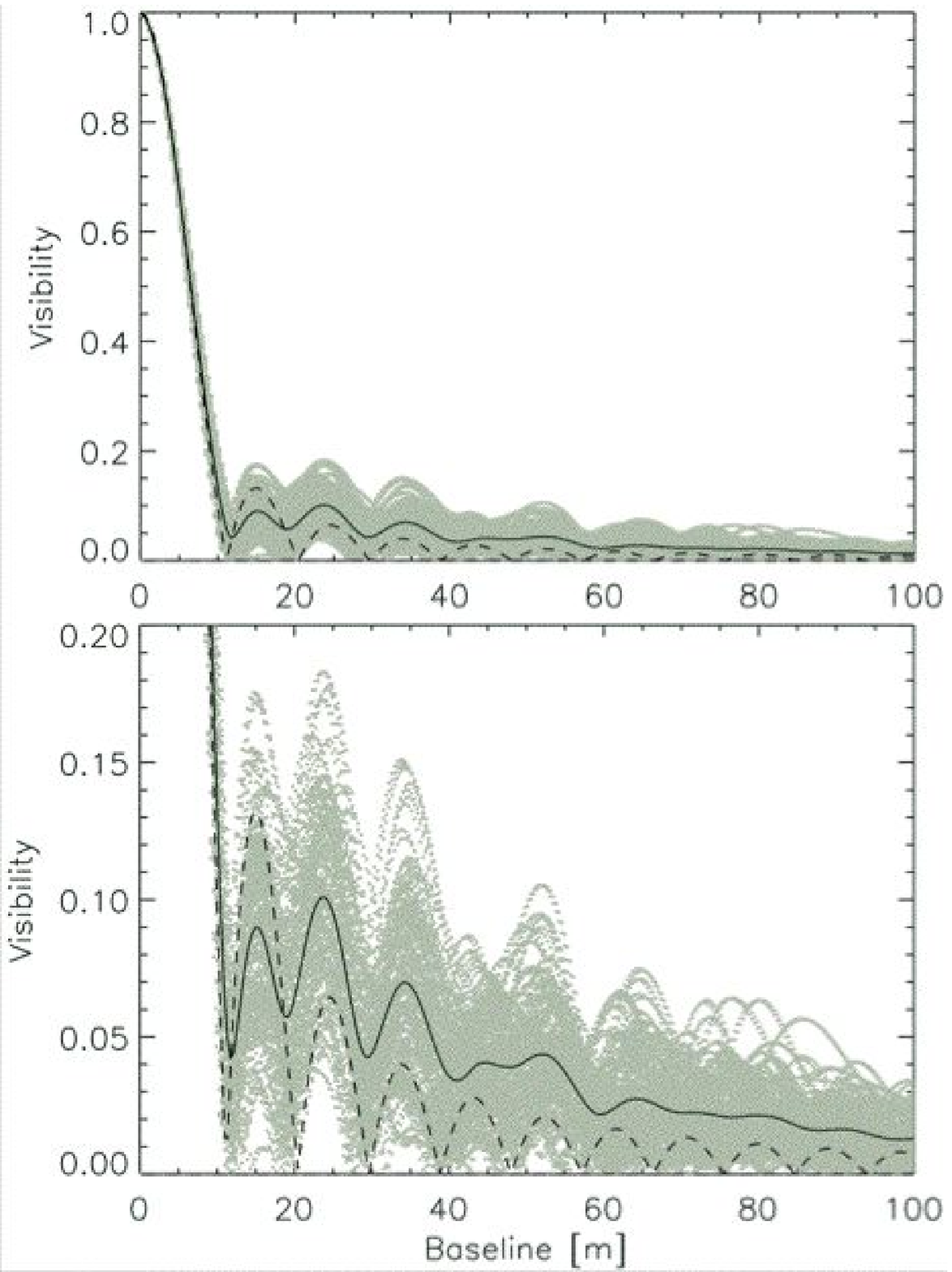} 
\end{tabular}
\caption{Scatter-plot of visibilities (grey) as a function of projected baseline length orientation, and the average profile (black), for $48$ snapshots and rotated images computed from the RHD simulation described in the text. The dashed curve is the visibility of a uniform disk with a diameter of $35$mas.  \emph{Left column} :  continuum at $1.6\mu m$. \emph{Right column} : Nearby CO line. Bottom panels are enlargements of the upper ones.}
\label{visibility}
\end{figure*}

We computed the visibilities curves from a large amount of intensity maps for different snapshots and rotated images (this analysis is extendable to other models and wavelengths). The visibilities carry a clear signature of deviations from circular symmetry (Fig. \ref{visibility}) because the small scale structures lead to a significant scatter. The emerging intensity at a given wavelength at a given position on the star depends on the opacity run through the atmosphere below that point. This in turn depends on the temperature and density stratification. Observations at a wavelength in a spectral line and in the nearby continuum probe thus different depth and thus different temperatures in the photosphere. In order to characterize the convective pattern, high spectral resolution is needed to determine the variations of the interferometric data between different spectral features, and between features and continuum.\\
Furthermore, the visibility strength at $100$m is $0.02-0.03$, much higher than what we expect for a uniform disk distribution. This is due to the small wiggle structures visible in the intensity maps (Fig. \ref{intensity}) that are not spatially resolved. \\
The convective-related surface structures are then observable with today's interferometers.

\subsection{Radiative-hydrodynamic models and Interferometry}

RHD simulations are necessary for a proper, qualitative and quantitative analysis of the surface of cool stars, in order to have parameter-free estimates of convection. RHD models are \emph{ab initio}, time-dependant, multi-dimensional, non-local and they contain as much physics as possible. The actual limitation is the CPU-time. RHD models are a great improvement over parametric models for the interpretation of the observations but they have to be validated with other techniques such as spectrometry and spectrophotometry (\cite{2006sf2a.conf..455C}). \\
In particular, VSI\footnote{VLTi Spectro-Imager is proposed as second generation VLTI instrument providing the ESO community with the capability of performing image synthesis at milli-arcsecond angular resolution} will produce reference observations for RSG stars from where the typical size of surface inhomogeneities, their contrast (in temperature and flux) and the order of magnitude for the time-scales of variations can be derived. These quantities are relevant for validating present models and for pushing further model developments.

\subsection{Future developments of the radiation transfer code}

The main characteristics of the 3D radiative transfer code we developed have been reported in Sect. \ref{3dcode}. On a short time-scale, future developments consist in introducing scattering and in achieving a 3D polarized radiative transfer code. This is particularly important to properly characterize the convective motions of RSG by constraining the degree of asymmetries and associated temperature contrasts. For this purpose, we plan spectropolarimetric observations. On a long time-scale, Non-LTE radiative transfer is required in order to properly analyse RSG spectra, since our final objective includes reliable determination of abundances for these stars. 

\section{Conclusion}

The comparison of our RSG RHD models with interferometric and spectroscopic observations is now possible with our radiative transfer code. \\
We found that convection-related surface structures are indeed observable and the interferometric visibilities carry an evident signature of deviation from circular symmetry. The phase and related closure phase show clear deviations from the axisymmetric case as well. The statistical approach applied in this work will be used to set a range for visibility/phase variations due to the photospheric convection to predict further observations (Chiavassa et al., in preparation).\\
There are few available interferometric observations suitable to constrain the models in terms of visibility and phase. Nevertheless, since RSG are prime targets for Interferometry, thanks to their large diameter and to their high-peak infrared luminosity, a lot of observations will be available in the near future. In this context, we think that a theoretical and observational simultaneous approach is really important to understand the atmospheric dynamics.

\begin{acknowledgements}
We thank Bertrand Plez and Eric Josselin for the useful and enlightening discussions and Bernd Freytag for the RHD models. We thank the "Centre Informatique National de l'Enseignement Sup\'erieur" (CINES) for 
providing us with computational resources required for part of this work. 
\end{acknowledgements}

%%-----------------------------
%%      your bibliography
%%-----------------------------

\end{document}